# FROM DIGITAL DIVIDE TO DIGITAL JUSTICE IN THE GLOBAL SOUTH: CONCEPTUALISING ADVERSE DIGITAL INCORPORATION

Richard Heeks, Centre for Digital Development, Global Development Institute, University of Manchester, UK, richard.heeks@manchester.ac.uk

**Abstract:** The connection between digital and inequality has traditionally been understood in terms of the digital divide or of forms of digital inequality whose core conceptualisation is exclusion. This paper argues that, as the global South moves into a digital development paradigm of growing breadth and depth of digital engagement, an exclusion worldview is no longer sufficient. Drawing from ideas in the development studies literature on chronic poverty, the paper argues the need for a new concept: "adverse digital incorporation", meaning inclusion in a digital system that enables a more-advantaged group to extract disproportionate value from the work or resources of another, less-advantaged group. This explains why inequality persists – even grows – in a digital development paradigm. To help ground future research and practice on this issue, the paper inductively builds a conceptual model of adverse digital incorporation with three main component sets: the processes, the drivers, and the causes of adverse digital incorporation. The paper concludes with thoughts on a future research and practice agenda that seeks to deliver digital justice in the global South: a necessary reconfiguration of the broader components of power that currently shape the inclusionary connection between digital and inequality.

## 1. INTRODUCTION

Inequality is one of the major challenges facing the world and there are significant concerns about the contribution of digital technology to inequality (UN 2020). The dominant lens for understanding the relation between digital and inequality has historically been that of the digital divide: of nations, regions, groups, individuals, etc. absolutely or relatively excluded from the benefits of digital technology (van Dijk 2020).

But we are now said to be moving towards a new phase or paradigm of the relation between digital and international development: "ICT4D 3.0" or "digital development" in which digital changes from being "a specific development tool to a general development platform" (Heeks 2020). What are the implications of this much broader, deeper role of digital in development for our understanding of the relation between digital and inequality?

Drawing from the development studies literature on chronic poverty, this paper argues that an increasing cause of inequality in the global South is not exclusion from digital systems but "adverse digital incorporation": inclusion in a digital system that enables a more-advantaged group to extract disproportionate value from the work or resources of another, less-advantaged group. To understand this process, the main task of this paper is to build a conceptual model of adverse digital incorporation given no such model – nor, even, of adverse incorporation generally – yet exists.

It does this through inductive review of the literature on adverse incorporation and the illustration of key components of adverse incorporation through their application to digital systems – platforms particularly – in the global South. The paper concludes with a graphic representation of adverse digital incorporation and consideration of the implications of this as a basis for digital justice.

# 2. BACKGROUND

*"When it is all said and done the telephone is not an affair of the millions. It is a convenience for the well-to-do and a trade appliance for persons who can very well afford to pay for it."* (The Times of 1902 cited in Mann 2010)

The telephone at the turn of the twentieth century may not have been a digital device but the quote above prefigures the concerns that arose with the growing diffusion of information and communication technologies (ICTs) during the last century. These crystallised in the mid-1990s with advent and growing use of the term "digital divide"; an idea which was soon applied on a global scale and became part of the parlance of international development (James 2003).

The digital divide was initially understood in Manichean terms: a dualism of "haves" vs. "have nots" that related to technology access; be it devices like PCs or services like Internet connectivity. Over time, the notion of the digital divide evolved and expanded in at least two ways (Ragnedda & Muschert 2013, van Dijk 2020):

   a) Forwards along the information value chain: particularly expanding from technology access to consider divides in technology use – for example, deriving from differences in user skills and knowledge.
   b) Backwards and outwards from the information value chain: particularly encompassing social inequalities (gender, race, disability, income, etc) that were seen as precursors to or even causes of the digital divide.

This broader and more contextualised view of the digital divide was sufficiently different from its origins that some sought to attach new labels, such as "digital inequality" (Robinson et al 2015, van Deursen 2017). Whichever the terminology, however, the foundational concept was exclusion and the underlying narrative was that particular groups or geographies were being prevented from accessing the benefits of digital technologies.

For those focusing on ICTs and development, such a worldview was generally sustainable in the first years of the twenty-first century as the majority of those in the global South were unable to access or use mobile phones or the Internet. The worldview remains relevant today to the hundreds of millions still without a mobile phone, the nearly three billion estimated to not use the Internet (ITU 2020), and all those unable to benefit from advanced digital applications like robotics or artificial intelligence. However, the worldview of exclusion is challenged in a world in which a significant majority of the global South's population have a mobile phone, and a majority have Internet access (*ibid.*). They are now included in, not excluded from, digital systems[1].

This tide of change – not just growth in access to digital infrastructure in the global South but far greater levels of usage and increasing depth of digitalisation and platformisation – is such that we have come to talk of a new "digital development" or "ICT4D 3.0" paradigm (Heeks 2020). One could still argue for the singular relevance of exclusion within this emerging phase of ICTs and development if the relationship between digital and inequality could be shown to relate solely to the declining category of those excluded from use of digital systems. Yet evidence suggests this is not the case; that, instead, inequality is increasingly related to use of digital technologies in the global South (Murphy & Carmody 2015, Gurumurthy et al 2019, Heeks & Shekhar 2021). Our understanding of digital and inequality must therefore encompass not just problems of exclusion but also problems of inclusion.

This shifting perspective mirrors earlier debates around poverty and development. Initial views saw poverty from the perspective of social exclusion: "the process through which individuals or groups are wholly or partially excluded from the society in which they live" (Hickey & Du Toit 2007:2).

---

[1] Defined here as socio-technical systems of digital data, digital technology, people and tasks (data processing and presentation, decisions, transactions, learning) (adapted from Heeks 2006).

The economic prescription flowing from this was to integrate the poor into markets. Yet, poverty and inequality persisted following globalisation and marketisation of developing economies. As a result, a new perspective arose: that of adverse incorporation which argued that some groups could be differentially disadvantaged through their inclusion in markets, states and civil society (*ibid.*).

The concept of adverse incorporation has achieved a niche presence within development studies; its use perhaps constrained by the lack of any clear and systematic framework for its application. However, it seems a relevant foundation on which to base investigation into the relationship between inclusion in digital systems and inequality; what, for some, will represent adverse digital incorporation. This investigation now follows – an inductive and iterative exploration of key concepts from the adverse incorporation literature and their illustration from digital development case studies. The adverse incorporation literature is not extensive and the analysis was undertaken based principally on three seminal sources that provide core insights into the concept: Bracking (2003), Hickey & Du Toit (2007) and Phillips (2013).

From this, as already indicated, adverse digital incorporation can be defined as inclusion in a digital system that enables a more-advantaged group to extract disproportionate value from the work or resources of another, less-advantaged group (adapted from Phillips 2013). Initial high-level thematic analysis of this adverse incorporation literature was also undertaken, which identified three core conceptual categories: systemic processes of unequal incorporation, drivers to incorporation, and causes of adverse incorporation. Each of these will be analysed in turn as elements in the development of a conceptual framework.

## 3. ADVERSE DIGITAL INCORPORATION CONCEPTS

### 3.1 Process Patterns of Unequal Incorporation

Drawing from the definition, then central to adverse digital incorporation is *exploitation* in the sense of the extraction of value by one group from the efforts of others (Phillips 2013). This can be seen at the level of individual workers and their labour. A digital development illustration would be the gig economy digital platforms that extract value from the labour of their workers, leaving too little value for the workers themselves[2]. Thus, for example, some of those working for gig economy platforms in South Africa find themselves earning below minimum wage and almost all find themselves earning below the living wage: the minimum amount deemed necessary to fulfil basic needs (Fairwork 2020a). Exploitation can also be seen at the level of enterprises. For example, small enterprises like hotels and travel agencies in Africa increasingly seek to participate in the global tourism markets run by digital platforms, in the hope of reaching direct to tourists particularly from the global North. However, the main beneficiaries are the platforms: "the promise of disintermediation remains unrealized for many as new kinds of foreign, internet-enabled intermediaries have emerged (e.g. TripAdvisor) to concentrate market power, control information about destinations, and achieve significant levels of capital accumulation outside Africa" (Murphy & Carmody 2015:203).

One part of this pattern of exploitation would be *commodification* in which something previously untraded is turned into a traded item; thus incorporating the owner or producer into a market. An extreme example would be the women – and children – from countries such as the Philippines who participate in webcam sex (Kuhlmann & Auren 2015, Mathews 2017). Their bodies are commodified for the benefit of Internet-connected others in distant places and, certainly in the case of children, often with long-term traumatic results for themselves.

Also related is *criminal exploitation* where individuals are drawn into participation in online activities in which value and resources are illegally extracted from them. While the stereotype of one type of this – "419ers", "Sakawa Boys" – has focused on those in global South countries as the perpetrators of fraud, they are also the victims. The Wangiri phone scam – after one ring, the call is cut and, when the recipient calls back, they are connected to a very high-cost premium-rate international call – has snared victims in Indonesia, Kenya, Pakistan and many other low-/middle-income locations (Priezkalns 2020).

---

[2] This relates to paid labour but exploitation can also be seen with unpaid labour. The most obvious example would be the actions of social media platforms which capture value from the unpaid labour of users sharing their thoughts, feelings, photographs, videos, etc.

These examples focus on extraction of value from labour or money but a further pattern of adverse digital incorporation is *legibility*: data about the existence or characteristics of a less-powerful group being captured in a digital system and thereby rendered visible to a more-powerful entity which then uses that data to enhance its power and control relative to the less-powerful group. Digital state surveillance systems throughout the world exhibit this pattern. China's Social Credit System is the current apotheosis of this, integrating data about citizens and their online behaviour from public and private digital systems that has "exponentially increased [*state*] capabilities to monitor the populace" (Liang et al. 2018:434) and constitutes a form of "data-driven authoritarianism" (Lee 2019:953).

While legibility's tropes of surveillance and control are particularly associated with the state, they are increasingly seen to affect workers in global South countries, as digital systems make them more legible to managers. Factory managers in China, for example, have required that workers all have mobile phones; that workers must respond immediately when called, even outside normal working hours, under threat of punishment for failure to do so; and that all messages are available for surveillance. As a result this digital device has become "a 'wireless leash' that shop-floor management can use as a nearly complete control and surveillance system over employees" (Qiu 2009:188).

If it were the digital equivalent of its physical predecessor, then *enclosure* would refer to the transfer into a privately- or state-owned digital system of what had previously been communal data or knowledge assets[3]. Misappropriation of traditional community knowledge relating to plants and animals – so-called "biopiracy" – is a relatively well-known example. For instance, a USAID-funded project captured from the Shuar indigenous community in Ecuador the details of hundreds of local plants and their medicinal uses (Nagan et al. 2010). This was then passed on to the US government National Cancer Institute which placed this knowledge into a closed-access information system for use by large pharmaceutical companies.

## 3.2   Structural Components of Adverse Digital Incorporation

*Drivers to Incorporation*

Why do individuals join digital systems that have adverse consequences for them?

In some cases, this arises from *ignorance*: a lack of knowledge of those adverse consequences and a belief that incorporation will be beneficial. We can see this at work in criminal exploitation. Scams targeting South Asian victims use Middle-East country codes; deceiving the recipients into thinking they have a call from relatives working in those countries (Javaid 2020). In this example, there is no benefit from incorporation into the scam but in other examples of adverse digital incorporation, the ignorance is more nuanced: the benefits do exist, even if not quite in the form or to the extent anticipated, and the ignorance is either of the existence of adverse consequences or of their likelihood and extent. For example, gig workers in Africa join digital platforms in the expectation of certain levels of income and without a clear understanding of the risks involved (Anwar & Graham 2021).

This highlights an important point about adverse digital incorporation: it may well not be solely adverse; i.e. solely negative in its consequences. Those participating in China's Social Credit

---

[3] Digital enclosure has also been rather more broadly applied to the capture of individual data (Andrejevic 2008). As with exploitation of unpaid labour, the most obvious example would be the actions of social media platforms which gain licence to distribute and use the thoughts, feelings, photographs, videos, etc. of individual users.

System receive benefits "such as deposit-free sharing economy services, fast-tracked check-ins for hotels, and mobile payment options" (Kostka 2018). Those working on gig economy platforms receive some level of income and some form of livelihood (Fairwork 2020a). Lack of benefit is not the essence of adverse digital incorporation. The essence, as noted above, is differential disadvantage – that a more-advantaged group disproportionately extracts value from the digitally-mediated actions or resources of the less-advantaged group; denying that latter group the value that should accrue to it and thus increasing relative inequality.

In other cases, the driver to joining an adverse digital system is ***direct compulsion***: a requirement of powerful others to join. Many state surveillance systems would fall into this category, for example where linked to a digitally-mediated identity that is then required in order to access public services. In India, there has been much criticism of the national identity database, Aadhaar, including its role as a state surveillance tool, and a capture of benefits by private interests, both legitimate and – in the case of privacy breaches – illegitimate (Khera 2019a). But "what started as a voluntary ID gradually became compulsory" (Khera 2019b:4): a "coercive application" of digital technology that leaves citizens with no choice but to participate (Basu & Malik 2017).

Neither of these drivers, though, satisfactorily explains many examples of adverse digital incorporation into economic digital systems where individuals join because of a lack of choice: an *exclusion* from better alternatives. Why is it, for example, that migrant workers join gig economy platforms even though they may well earn less than minimum wage? In the case of South Africa, a number of those interviewed for the Fairwork project did so because they were excluded from other employment opportunities either by legal requirements or lack of social capital or by discriminatory hiring norms. Likewise for parents prostituting their children online in the Philippines, "the first factor is poverty … they tend to engage in that so that they can have enough food to eat" (Kuhlmann & Auren 2015:38). These families live physically, socially and economically on the margins of cities and they are excluded from systems of formal employment and welfare.

Exclusion and adverse incorporation are thus not mutually-exclusive perspectives in understanding digital inequality but can be closely connected (see also Hickey & Du Toit 2007). Historical and contextual patterns of exclusion from particular economic, social and political systems can significantly increase the likelihood that marginalised individuals and groups will participate in digital systems that are disadvantageous. Any understanding of adverse digital incorporation must therefore encompass ***temporality*** and ***contextuality*** – the historical and contextual processes by which those incorporated have come to be excluded from alternative systems.

*Causes of Exploitation*

Once incorporated into a digital system, why is it that the value of actions and/or resources is differentially distributed? The literature on adverse incorporation is repetitively clear that the root cause for this is power and control: the way in which a more-advantaged group controls the system into which the less-advantaged group is incorporated (Bracking 2003, Hickey & Du Toit 2007). That control allows the former to extract and capture the value generated by the latter.

In a very direct sense, inequitable outcomes emerge from digital systems because the more-advantaged group has control of design of the system: a ***design inequality*** compared to the exploited users. This was the case in almost all of the instances given above: that states or platform companies are able to design the processes and governance of digital systems in such a way that resources flow unequally. This is often most visible when alternative designs exist which indicate there is nothing inherent in the inequalities that are found. For example, some mapping systems are extractive: using outsiders to take data from low-income communities and then present it online for the use and benefit of others. But alongside such designs are participative others planned by or with the community. These use community members to undertake the mapping, and make specific efforts – through low-tech interfaces, paper-based maps, presentations at community meetings – to enable communities to make use of that data (Heeks & Shekhar 2019).

*Resource inequality* can lie behind unequal outcomes of using digital systems. Users with lower access to financial, human, social, physical and informational capital will be differentially incorporated into digital systems compared to those with higher endowments. In the biopiracy case, for instance, it is the global North actors who know the economic value of local plants when the Shuar do not, and it is the former who have financial resources, socio-economic contacts and physical machinery necessary to monetise the plants into pharmaceutical products (Nagan & Hammer 2013).

*Institutional inequality* can play a role, where formal laws and regulation and informal norms and values favour the more-advantaged group. For example, East African small enterprises digitally integrating into global value chains often struggle; suffering greater volatility and risk with the potential for profits to be reduced (Foster et al 2018). The beneficiaries are the lead firms in the global supply chains; those which determine the specifications and standards that African small enterprise must adhere to, and which use the flows of digital information to more tightly-control their suppliers and to switch from less- to more-adherent suppliers.

*Relational inequality* can be understood in terms of the relative dependencies between the actors within a digital system. In the economic sphere, the substantial reserve army of labour in many South countries creates asymmetrical dependency. For example, physical gig platforms employing drivers and deliverers can readily replace any individual worker (Gomez-Morantes et al 2019). The platforms therefore do not depend on the worker and are able to treat them adversely. On the other side, individual workers may depend significantly on the platform; particularly if – based on the expectation of a certain, stable income – they have taken out loans. It has been shown that, the greater the dependency of the worker on the platform, the more willing they are to allow themselves to be exploited (Schor et al. 2020). The asymmetry of dependency in this relationship is exacerbated by the atomisation of gig workers in the general absence of trade unions or worker associations (Graham et al 2017). The structural relationship of platform to workers is thus many individual one-to-one relations rather than a one-to-many relation mediated by a worker association; the former being considerably weaker and more open to exploitation.

### 3.3 Conceptual Framework

Having drawn out the key components of adverse digital incorporation, we can put them together

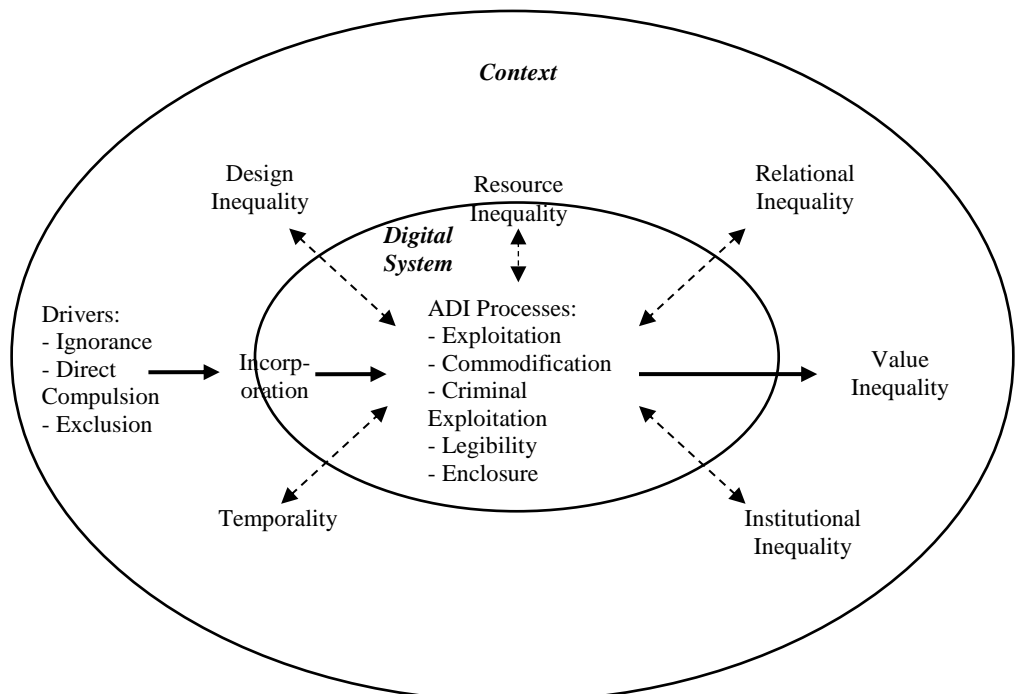

**Figure 1. Conceptual Framework of Adverse Digital Incorporation**

Source: author

into a single overall conceptual framework, as shown in Figure 1. This centres the digital system and the patterns of adverse digital incorporation described in the first sub-section above. Drivers to incorporation are identified on the left and the main outcome of unequal extraction of value is identified on the right side. Around the system are the other structural components that facilitate adverse digital incorporation, including the historical perspective of temporality.

## 4.   CONCLUSIONS

The era of digital development will be marked by many developmental benefits of digital systems. Equally, the problems associated with such systems will not be confined to inequality: the growing carbon footprint of digital systems is but one example. However, digitally-related inequality is likely to be a major challenge throughout the century for those involved in digital development.

The concept of digital divide and related ideas such as digital inequality have moved beyond their initial simplistic origins. However, they have to date remained rooted in a worldview of exclusion from the benefits of digital systems. In this paper I argue that, as we move into a digital development paradigm, this worldview remains important but it is no longer sufficient. We need as well to account for the inequalities that arise in the global South when less-advantaged individuals and groups are included in rather than simply excluded from digital systems. Drawing from development studies, this paper argues that the concept of "adverse digital incorporation" can help to understand the emerging relation between digital and inequality. To help operationalise this new concept, the paper then built a model of it.

Having created the adverse digital incorporation model inductively, a next step for digital development research will be to apply the model deductively as the basis for analysing case studies in which use of digital systems is associated with unequal outcomes. Case examples have already been suggested above but others are likely to increasingly emerge. In terms of research paradigm, the emphasis on causal mechanisms linking structural precursors to processes of exploitation suggests that critical realism may be an appropriate frame. Methodologically, and given the need to understand context, relations, differential extraction of value, etc., then qualitative methods are likely to be of most relevance for such research.

The era of digital development seems likely to be being marked with a growth in adverse digital incorporation. Is this simply a reproduction of existing processes of adverse incorporation in a digital milieu? Or is there something inherent in the functionalities and affordances of digital systems that makes them more likely to facilitate, or even to create, unequal outcomes? Put another way, where is the <u>digital</u> in adverse digital incorporation: is Figure 1 in fact just a model of adverse incorporation?

Answering such questions must be part of the future research agenda but we can start to identify some of the paths for exploration of this issue. Research on the institutional work of digital platforms suggests that they enable an aggregation of market institutional functions that were previously distributed and dissipated (Heeks et al 2021). This enables an aggregation of power well beyond that feasible in traditional markets, and hence a greater asymmetry of power between platform owner and platform users. In turn this greater asymmetry enables a disproportionate extraction of value. Digital has also – via machine learning and algorithms – made systemic processes such as decision-making or distribution of value more opaque (Burrell 2016). Such opacity hides and thus facilitates disproportionate extraction of value.

The focus here has been on the victims of adverse digital incorporation but research will also be needed on the beneficiaries. What drives them to design and implement exploitative digital systems? Can we find some systematic difference between those creating systems that increase inequality and those creating systems that decrease inequality?

This last question moves us into the realm of practice. In practical terms, countering adverse digital incorporation would mean identifying digital systems that unequally include already-disadvantaged groups and seeking to address the drivers, causes or processes of adverse digital incorporation. An example here would be the Fairwork project, which seeks to address inequalities between capital and labour that emerge as gig workers are adversely incorporated into digital labour platforms (Fairwork 2020b). It does this by intervening in resource inequality – providing workers with open information about pay and conditions on platforms; and by intervening in institutional inequality – encouraging standards and norms for decent gig work and for ethical consumption and investment in the gig economy.

A danger of contextual models such as the one developed here is that they lapse into structural determinism: assuming that only external structural interventions can improve the impact of adverse digital systems, and failing to recognise the agency of those who have been adversely incorporated. Taking again the example of gig platform workers, we can see many examples around the world of them self-organising and taking protest or legal action to reduce the unequal distribution of value that derives from their labour (e.g. Wood et al 2018, Joyce et al 2020). The potential for agency of disadvantaged groups must therefore be part of the agenda for development practice.

Recognition and conceptualisation of adverse digital incorporation offers a basis for alternative digital development design strategies. "Neutral digital incorporation" would design digital systems in which value was evenly rather than unevenly distributed between system actors. "Advantageous digital incorporation" would design digital interventions that specifically sought to reduce existing inequalities. Based on the understanding developed above, the key insight is that advantageous digital incorporation can only occur if digital interventions in some way address underlying inequalities, both historical and contextual.

As summarised in Figure 2, a key argument in the domain of inequality is that only justice – rather than equality or equity – will truly address inequality in the long term because it addresses the underlying causes of that inequality rather than just dealing with its manifestations. Using this terminology suggests we must therefore move beyond the digital divide, and beyond digital inequality, to "digital justice": seeking to address not just the proximal processes of adverse digital incorporation but also their underlying causes.

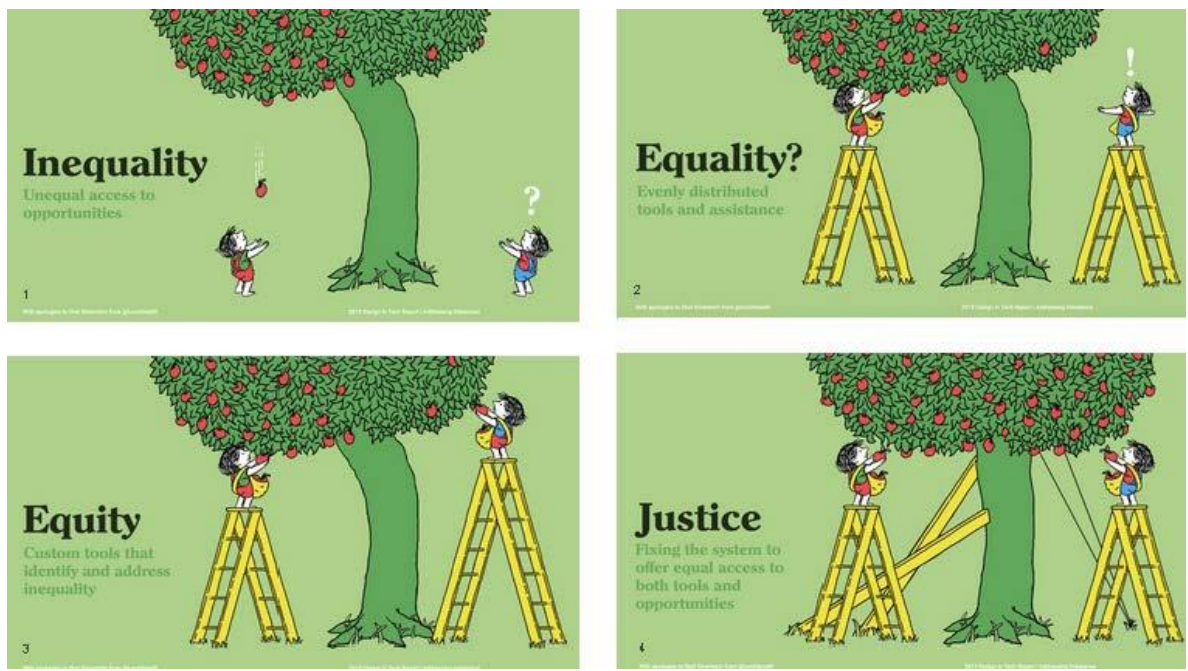

**Figure 2. Equality, Equity and Justice**

Source: Tony Ruth from Maeda (2019)

As the Figure 1 model demonstrates, this takes the focus away from the practices and procedures of digital development systems and towards the need to impact the wider institutions, structural relations, design processes and resource distributions that surround such systems. Only by impacting those can we move from adverse to advantageous digital incorporation, and deliver digital justice in the global South.

**Acknowledgements**

This paper builds on and uses material from "From the Digital Divide to Digital Justice in the Global South", a keynote presentation made at the workshop on "Digital (In)Equality, Digital Inclusion, Digital Humanism" held as part of the 12th ACM Web Science Conference 2020. The insights of IFIP WG9.4 conference reviewers are also acknowledged.